\documentclass[cameraready]{Interspeech}

\usepackage{amsmath,amssymb,amsfonts}
\usepackage{booktabs}
\usepackage{multirow}
\usepackage{microtype}
\usepackage{float}

\title{A Head-Level Investigative Study of Self-Supervised Backbones for Synthetic Speech Detection}

\author[affiliation={1}]{Mohamed}{Mallat}
\author[affiliation={1}]{Michele}{Panariello}
\author[affiliation={1}]{Massimiliano}{Todisco}
\author[affiliation={2}]{Anthony}{Larcher}
\author[affiliation={1}]{Nicholas}{Evans}
\address{$^1$ EURECOM, Sophia Antipolis, France \\
         $^2$ LIUM, Le Mans Universit\'e, France}
\email{mohamed.mallat@eurecom.fr, michele.panariello@eurecom.fr, massimiliano.todisco@eurecom.fr, anthony.larcher@univ-lemans.fr,
evans@eurecom.fr}


\begin{document}
\maketitle

\vspace{-1cm}
\begin{abstract}
Self-supervised learning (SSL) speech backbones are the standard frontend
for synthetic-speech detection, and layer-wise probing has established which
layers carry the most signal. But analysis stops there: which of a layer's
twelve attention heads compute the spoof decision is unknown. We present the
first head-level study of two frozen backbones, WavLM-base and Wav2Vec2-base,
on ASVspoof~5. Per-head ablation shows that only a few heads carry the
decision while others actively oppose it; removing those heads lowers equal
error rate (EER) with no retraining, by 13\% relative in WavLM and 22\% in
Wav2Vec2. We also find a probe-free signature: spoofed speech draws more
focused attention than genuine speech, and this gap alone separates the two
almost as well as the full representation, peaking at each backbone's most
discriminative layer. Finally, attack identity is decodable from every head,
yet heads differ in what they do, not in what they know.
\end{abstract}

\section{Introduction}
\label{sec:intro}

Layer-wise probing studies have shown that the intermediate transformer
layers of self-supervised learning (SSL) speech models such as
WavLM~\cite{chen2022wavlm} and Wav2Vec2~\cite{baevski2020wav2vec2} often carry the most information to distinguish bona fide from spoofed utterances~\cite{iwbf2026wavlm,naacl2025layerwise,wang2024wavlm}. 
Such studies, though, do not reveal \emph{what inside} the layer contributes most.
A transformer layer is not one unit but a set of attention heads~\cite{vaswani2017attention} (usually twelve), each with its own attention pattern and output.

While typically, the set of heads are treated as a single, indivisible block, their inner workings are not unexplored territory. 
Head ablation and head-role taxonomies have been established for natural language processing~\cite{michel2019sixteenheads,voita2019heads}. 
For SSL speech
models, prior head-level work has explored head pruning to reduce model complexity while not degrading benchmark scores~\cite{wang2023taskagnostic}.
Speech transformer attention patterns have also
been explored, including through
entropy~\cite{sat2020understanding,crossling2024attention}, to characterize how SSL models attend to generic speech. 
To the best of our knowledge, no published work has examined the head attention patterns of SSL-based spoof detection backbones and explored what each contributes to spoof detection. 
SSL spoof detection heads operate in cooperation.
However, while some heads contribute constructively, others actively degrade detection reliability. 
While the model is often successful in detecting spoofed inputs, particularly spoofed utterances similar to those seen in training, detection often fails when the model encounters differently-generated spoofed utterances seen in evaluation data.  
This is the classic \emph{generalization problem}~\cite{muller2022does} for which, as we will show, a sub-set of heads can be responsible.

We study two frozen backbones, WavLM-base and Wav2Vec2-base, using the ASVspoof~5 corpus~\cite{wang2024asvspoof5}.
We train lightweight probes for each layer and we pick the most discriminative one within which we drop one head at a time and measure what each adds to detection using a set of spoof attacks not seen during training. 
Our research questions are as follows.
Which heads matter?
How few are enough? 
What does each head reveal? 
We find that the answer is
not only in the representations the probe reads.
Spoofed utterances leave a visible mark on the attention patterns themselves.

Our contributions are as follows: 
\textbf{(i)} we show the spoof decision is carried by only a sparse subset of the attention heads;
\textbf{(ii)} a subset of heads degrades detection and their removal improves detection performance without retraining;
\textbf{(iii)} improvements are concentrated on the most difficult to detect attacks;
\textbf{(iv)} spoofed speech carries a probe-independent attention-entropy signature that tracks discriminability across the layer stack and peaks at each backbone's most discriminative layer, independently of the attack set used to measure it;
\textbf{(v)} this signature is informative on its own, detecting unseen attacks almost as well as a probe over the full representations; 
\textbf{(vi)} attack identity is decodable from every head, yet the heads differ in what they do, not in what they know.

\begin{figure*}[t]
  \centering
  \includegraphics[width=0.70\textwidth]{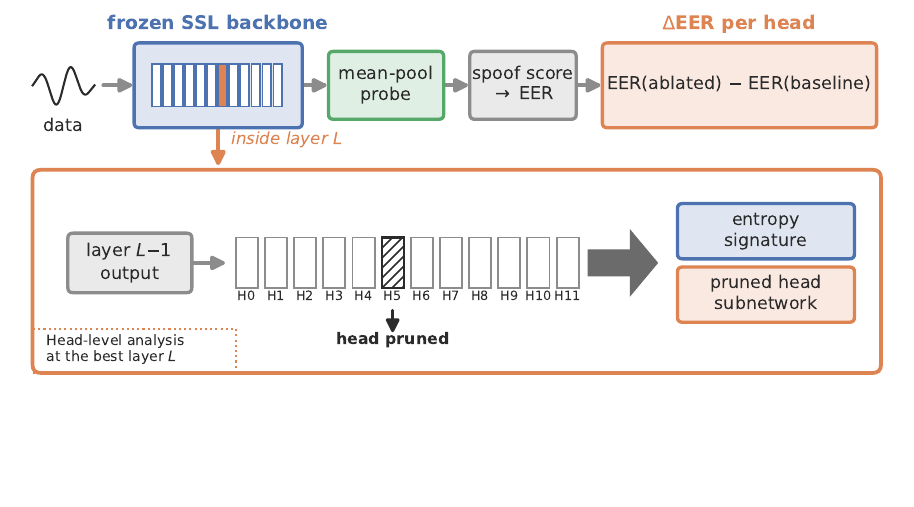}
\vspace{-0.8cm}
  \caption{Overview. A frozen SSL backbone feeds a mean-pool probe; at the
  best probe layer we remove one attention head at a time and read the
  change in EER, which sorts the twelve heads into detectors (removal
  hurts), interfering heads (removal helps) and redundant heads. Two readouts
  follow: a probe-free attention-entropy signature, and a pruned head
  subnetwork that lowers EER.}
  \label{fig:overview}
\end{figure*}
\vspace{-0.4cm}
\section{Related work}
\label{sec:relwork}

\textbf{SSL backbones for spoof detection -} The use of fine-tuned or frozen SSL frontends is the dominant approach to spoofed speech detection~\cite{tak2022sslantispoofing,jung2022aasist}.
A growing body of work has explored where in the backbone the most useful information is extracted and recent layer-wise probing and learned layer weighting studies~\cite{pasad2021layerwise,iwbf2026wavlm,naacl2025layerwise,wang2024wavlm} point consistently to intermediate layers. 
All of these studies treat the layer as the main unit of analysis.

\noindent\textbf{Attention of speech transformers -} Each attention head produces an attention map, a matrix of where it attends across the utterance; prior work categorizes these per-head maps by shape, into local (diagonal), reference-point (vertical), and diffuse (global). Entropy is the standard tool for quantifying how focused or spread a head's attention is~\cite{sat2020understanding,crossling2024attention,clark2019what}. 
While the analysis reveals where a head attends, the work has considered generic speech rather than attention behavior for spoof detection.

\noindent\textbf{Head-level interventions -} Head ablation has been studied widely in natural language processing and used to explore
their functional roles~\cite{michel2019sixteenheads,voita2019heads}. For SSL speech models, head pruning has been used to reduce model complexity~\cite{wang2023taskagnostic}, 
the output is, however, only a binary mask and is far from a real understanding of functional role. 
We instead use head ablation as a causal probe: rather than describing where a head attends, we remove the head and measure the change in detection error, revealing whether it helps, hurts, or does not affect the spoof decision.

\vspace{-0.3cm}
\section{Probing and head-level analysis}
\label{sec:method}

Figure~\ref{fig:overview} illustrates the analysis pipeline. We use frozen SSL backbones and train lightweight layer-wise probes to identify the most discriminative transformer layer. We then perform head-level interventions at that layer, using single-head ablation and head-subset sufficiency tests to determine which heads help, hurt, or have little effect on spoof detection. Finally, we analyze attention statistics and per-head functional roles to separate what information each head carries from how it contributes causally to the full detector.

\vspace{-0.15cm}
\subsection{Dataset and probing protocol}
\label{ssec:dataset-probing}
\vspace{-0.15cm}
We use the ASVspoof 5 data \cite{wang2024asvspoof5}, an English corpus whose spoofed utterances are generated by text-to-speech and voice-conversion systems (speech deepfakes). Probes are trained on the training split (A01–A08) and analyzed on the dev split (A09–A16), so every attack we analyze is unseen at training time. The evaluation split, which contains attacks A17--A32, is held out and used only once as a generalization check for head roles.

We use two frozen SSL backbones: WavLM-base and Wav2Vec2-base. For each transformer layer, we train a linear probe consisting of mean pooling over time followed by a two-class linear classifier. The layer with the lowest development-set EER is selected as the best probe layer upon which all head-level analysis is performed. 

We analyze a sample of 2000 bona fide utterances, selected evenly across speakers and 2000 utterances per attack, giving a total of 5200 utterances. All speech data are sampled at 16~kHz and cropped to 4 seconds duration. We rank the heads and choose the sufficiency subset on one random subsample of utterances, then report EER on two further, non-overlapping subsamples. This way a head is never both selected and scored on the same data, which would otherwise overstate the result.

\subsection{Head removal and sufficiency}
\label{ssec:head-removal}

At the selected layer $L$, an attention head $h$ is removed by zeroing its corresponding block of columns in the attention output projection,
\begin{equation}
W_O[\,:\,,\,h\,d_h:(h{+}1)\,d_h\,] \leftarrow \mathbf{0},
\end{equation}
where $d_h = 64$ is the head dimension. Since heads within a layer do not interact before the output projection, this operation removes the contribution of head $h$ to the layer output while leaving the rest of the backbone unchanged.

We first remove one head at a time and record the change in development EER relative to the full layer. This produces a per-head causal map: heads whose removal increases EER are interpreted as detector heads, heads whose removal decreases EER as interfering heads, and heads with negligible effect as redundant or weakly used heads. We then perform a sufficiency test which consists of ranking heads according to this map, keeping only the top-$k$ heads and zeroing the rest, for $k = 1,\ldots,12$.

\vspace{-0.1cm}
\subsection{Attention-entropy statistics}
\label{ssec:entropy}

For every head, we compute its \emph{attention entropy}, the per-frame entropy of its attention distribution averaged over the $T$ frames
\begin{equation}
\bar{H}
=
-\frac{1}{T}
\sum_{i=1}^{T}
\sum_{j=1}^{T}
a_{ij}\log_{2} a_{ij},
\end{equation}
where $a_{ij}$ is the attention weight from frame $i$ to frame $j$. Frames are the backbone's 20~ms output steps ($T=199$ for 4~s of audio), not spectrogram frames. Lower entropy indicates more focused attention.
This statistic depends only on the backbone and the input, not on the trained probe.

Collecting the entropy values from all 144 heads gives a fixed-length attention signature for each utterance. To test whether this signature carries spoofing information on its own, we train a logistic-regression classifier under two unseen-attack protocols. In the first, the classifier is trained using the training split and tested using the development split, where all eight attacks are unseen. In the second, we use leave-one-attack-out validation within the development split, holding out one attack in each fold. As a control, we also check whether attention mass correlates with frame energy, to rule out a simple loudness artifact.  To confirm the gap is real rather than noise, we test each head's entropy difference between bona fide and spoofed speech for significance, using a Wilcoxon rank-sum test with a Bonferroni correction for the 144 heads tested.

 \begin{figure}[t]
  \centering
  \includegraphics[width=0.88\columnwidth]{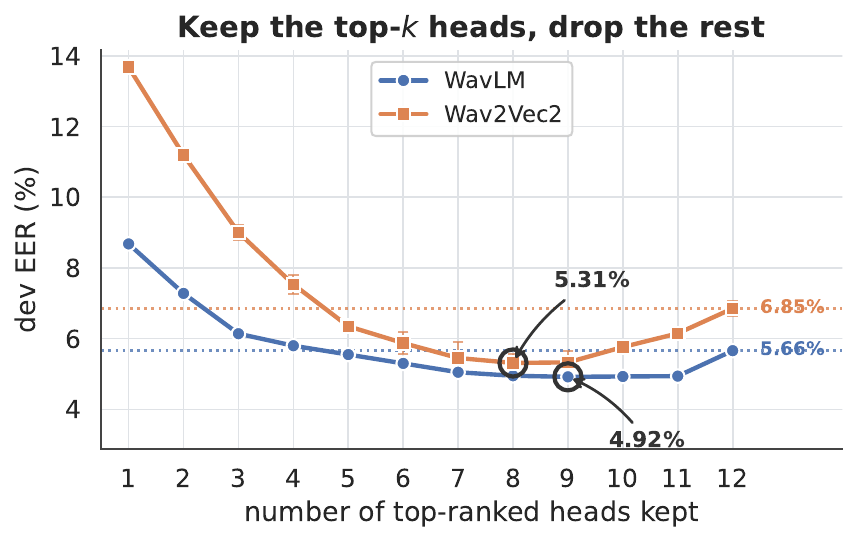}
  \caption{Sufficiency: dev EER when only the top-$k$ heads of the best
  layer are kept.}
  \label{fig:sufficiency}
\end{figure}

\begin{figure}[t]
  \centering
  \includegraphics[width=0.88\columnwidth]{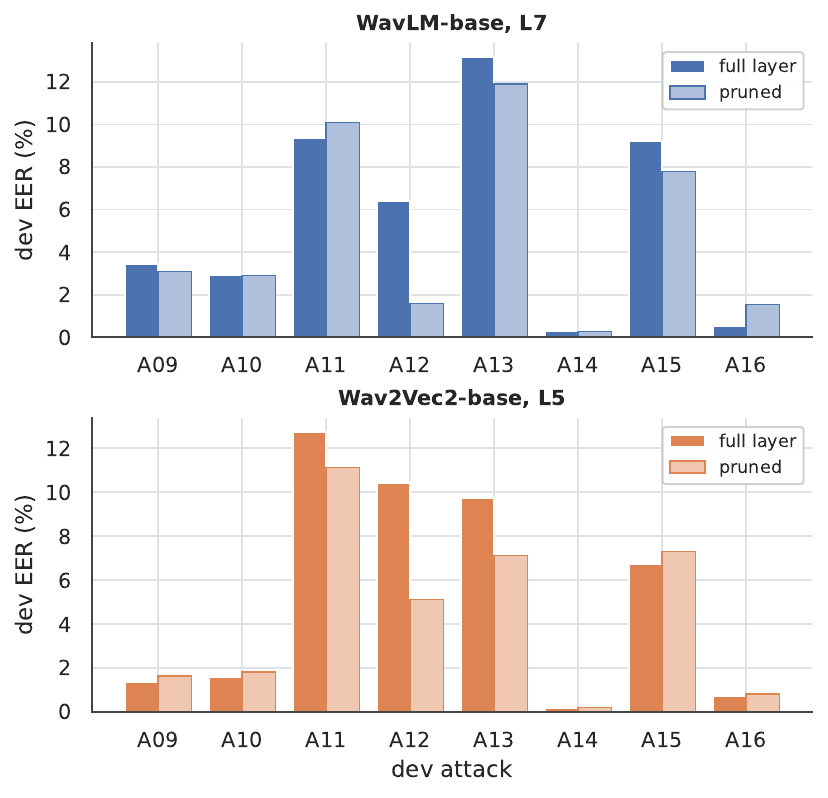}
  \caption{Per-attack dev EER before (full layer, solid) and after pruning
  to the best subnetwork (lighter).}
  \label{fig:perattack}
\end{figure}

\subsection{Information and causal views of head function}
\label{ssec:head-function}

We analyze head function from two complementary perspectives. The information view asks what is decodable from an individual head's output. For each head at the best layer, we average its 64-dimensional output over time and train two logistic-regression classifiers: a binary bona fide versus spoof classifier (trained on the training split, EER on the development split) and an eight-way classifier of the development attacks (spoofed utterances only, 4-fold cross-validated accuracy).

The causal analysis asks what head contributes to the full detector. This view uses no additional classifier: it reads the per-attack breakdown of the head-removal map and measures how removing the head changes the full-system EER. We call a head attack-conditional if its removal improves detection for some attacks while degrading it for others. This distinction allows us to separate what a head knows from what it does in the complete spoof-detection system.

\vspace{-0.3cm}
\section{Results}
\label{sec:results}

\subsection{The spoof signal is concentrated in a few heads}
\label{sec:res:concentration}
\vspace{-0.18cm}
Layer-wise probing selects L7 for WavLM (development-set EER 5.4\%) and L5 for Wav2Vec2 (6.4\%). We remove each head at the best layer, one at a time. For both backbones a handful of heads contribute most towards detection: for WavLM L7, the removal of H8, H0 or H6 raises EER by 1.5, 1.4 and 1.2 percentage points, respectively, while most of the remaining heads have negligible impact. Wav2Vec2 L5 shows the same concentration. The layer is not a uniform bag of redundant heads; a small subset does the work.

\subsection{A few heads suffice, and some actively interfere}
\label{sec:res:sufficiency}
\vspace{-0.1cm}
Figure~\ref{fig:sufficiency} shows results for when only the top-ranked heads are retained; others are removed (set to zero). Keeping a small subset of the twelve heads already matches or outperforms the full, un-pruned layer~\cite{frankle2019lottery} for both backbones and, for both, a part of the layer actively interferes. For WavLM, removing the three lowest-ranked heads in the best layer (keeping nine) lowers the EER from 5.66\% to 4.92\%, a 13\% relative reduction with no retraining. For Wav2Vec2, keeping eight heads lowers the EER from 6.85\% to 5.31\%, a 22\% relative reduction. For both backbones, 
pruned heads are the same as those whose ablation is beneficial to detection when removed one at a time. The interference is hence consistent with the per-head map rather than an artifact of one draw. Improvements are concentrated on the most challenging attacks in the development set (A12, A13, A11), namely those for which detection performance is worst for the full layer
(Figure~\ref{fig:perattack}). Interfering heads are thus a general property of the layer, and are not specific to one backbone.
\begin{figure}[t]
  \centering
  \includegraphics[width=\columnwidth]{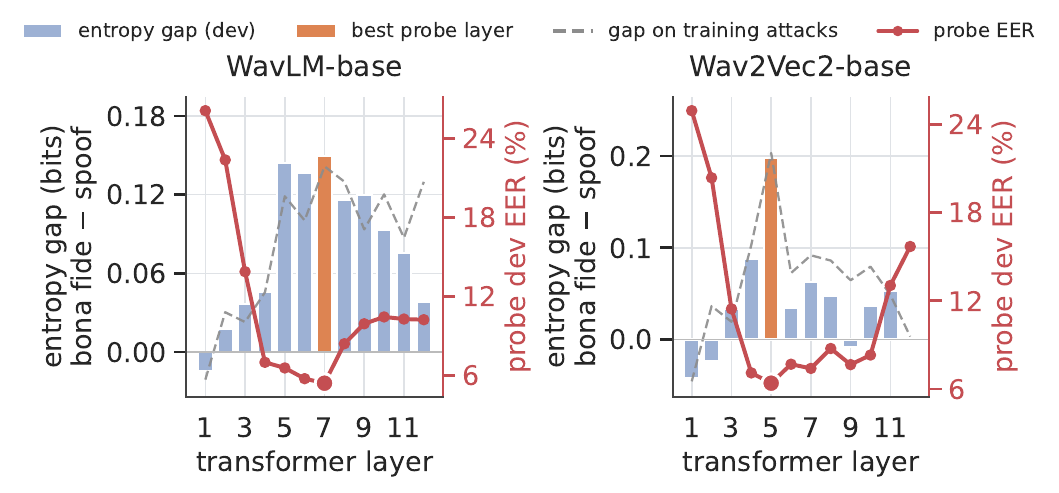}
  \caption{The attention-entropy signature. The
  gap tracks discriminability and peaks at each
  backbone's best probe layer (orange bar).}
  \label{fig:entropy}
\end{figure}
\subsection{Spoofed speech focuses attention}
\label{sec:res:entropy}
\vspace{-0.18cm}
All previous results are derived using a probe. The next set of results are not. For each head we
measure the entropy of its attention, a quantity that depends only on the backbone and the input. Spoofed utterances induce lower-entropy, more focused attention than bona fide utterances; the gap reaches up to 0.15 bits in WavLM and 0.20 bits in Wav2Vec2, and changes from layer to layer (Figure~\ref{fig:entropy}). Across the twelve layers, the per-layer
entropy gap and the per-layer probe EER are strongly inverse-correlated
(Spearman $\rho = -0.87$ and $-0.76$), and the gap peaks at exactly the
most discriminative layer of each backbone: L7 for WavLM, L5 for
Wav2Vec2. The peak corresponds consistently to the same layer when measured using the disjoint
training-attack set, so the peak reflects the backbone and not the particular attack. Per head, the difference is significant for 130 of 144 heads for
WavLM and 132 for Wav2Vec2 (Wilcoxon rank-sum, $p<0.05/144\approx3.5\times10^{-4}$). A natural concern is that focused attention merely tracks the highest energy frames, and that spoofed and bona fide audio differ in terms of energy. This is not the case: the attention of the best performing detector heads is uncorrelated with frame energy ($|r| \le 0.1$). A plausible explanation is that 	synthesis artifacts are localized in time: a few frames betray the generator, so the heads lock onto them, whereas natural speech offers no such anchor and attention stays spread.

\vspace{-0.1cm}
\subsection{The signature detects unseen attacks on its own}
\label{sec:res:detector}
If attention entropy distinguishes bona fide from spoofed speech, it should be sufficient on its own.
We summarize each utterance by the set of 144 per-head
entropies and fit a logistic regression (Table~\ref{tab:entropy_det}).
When trained using only attacks in the training set and tested using the eight unseen attacks in the development set,
we reach an EER of 5.95\% using WavLM, within 0.6 points of the
full-representation probe (5.4\%), and 9.69\% using Wav2Vec2. Leaving one attack out at a time gives a mean EER of roughly 3\%, lower than the 5.95\% above. In this setting each held-out attack is scored against the other development-set attacks, which belong to the same generation of spoofing systems; this is easier than the train-to-development jump, where the training and development attacks come from different generations. Attention sharpness alone, with no access to the representations, is a strong spoof cue.

\vspace{-0.2cm}
\subsection{Heads differ in what they do, not what they know}
\label{sec:res:dissoc}
\vspace{-0.1cm}
Finally we investigate what is decodable from an individual head. We apply the eight-way attack classifier of Section~\ref{ssec:head-function} to each L7 head. The attack is decodable from every head, at 76 to 82\% accuracy (chance 12.5\%), yet no single head is a reliable detector on its own: the best single-head binary classifier reaches 11.3\% EER, versus 5.4\% for the full layer. Information about the attack is thus everywhere, and a head's ability to generalize to an unseen attack is uncorrelated with how attack-specific its output looks (Table~\ref{tab:dissociation}). In other words, how much attack information a head carries says little about whether it actually helps the detector. Function is not uniform, however. Some heads help every attack (WavLM H8 is both the most general and the most damaging to remove), while others are
attack-conditional, assisting in the detection of some attacks while degrading the detection of others: for WavLM L7,
H5 works against detection on most attacks yet contributes to the detection of A11.
For Wav2Vec2, the four interfering heads of
Section~\ref{sec:res:sufficiency} are all attack-conditional. The spoof
decision is influenced by
heads with distinct functional roles, not by a group of interchangeable detectors.
\begin{figure}[t]
  \centering
  \includegraphics[width=0.82\columnwidth]{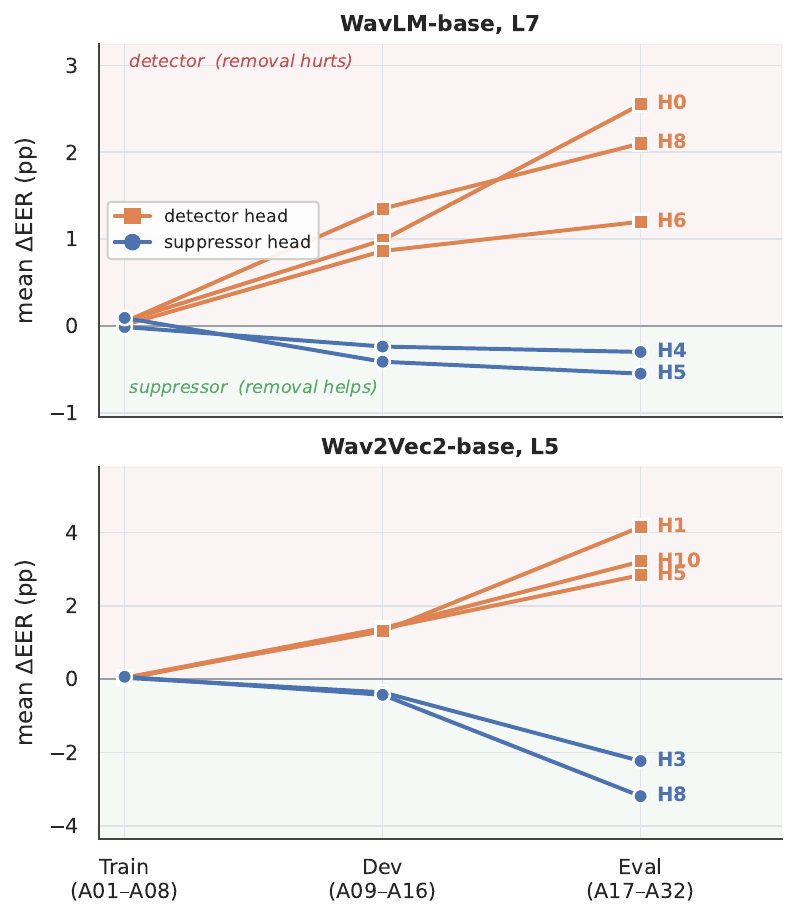}
  \caption{Head roles transfer to unseen attacks, in both backbones. Mean
  $\Delta$EER when a head is removed, averaged over the attacks in each
  group.}
  \label{fig:generalization}
\end{figure}

\begin{table}[t]
\centering\small
\caption{Detection from attention entropy alone (144 values per
utterance) under unseen-attack protocols, against
the mean-pool probe on full representations.}
\label{tab:entropy_det}
\resizebox{\columnwidth}{!}{%
\begin{tabular}{lcc}
\toprule
 & WavLM-base & Wav2Vec2-base \\
\midrule
Probe, full representations & 5.4\% & 6.4\% \\
Entropy only, training$\rightarrow$dev & 5.95\% & 9.69\% \\
Entropy only, LOAO (mean) & 2.98\% & 3.49\% \\
\bottomrule
\end{tabular}}
\end{table}

\vspace{-0.3cm}
\subsection{Head roles hold up on newer, unseen attacks}
\label{sec:res:generalization}
\vspace{-0.1cm}
The roles above are derived from detection performance for development set attacks. To check they are not an
artifact of that set, we re-measure the effect of removing each head using the
ASVspoof~5 evaluation set (containing A17--A32), a set of attacks unseen in training, and the training set (attacks A01--A08) for
reference. Figure~\ref{fig:generalization} shows the pattern is stable and,
if anything, sharper out of distribution. For training attacks the
probe is saturated (EER near zero) and every head looks inert; for the development set the head roles appear; for the evaluation set, detector heads matter more
(WavLM H0 and H8 grow from about $1.0$--$1.4$ to $2.1$--$2.6$~pp) and the
interfering heads keep degrading detection performance. Wav2Vec2 behaves the same way, with
interfering heads causing it to drop $-2$ to $-3$~pp on eval. The different head roles we observe for the development set are therefore stable across attack sets rather than an artifact of the development attacks, and become more pronounced. as the attacks drift further from those seen in
training.

\begin{table}[t]
\centering\small
\caption{WavLM L7 heads: information is uniform, function is not. A head is attack-conditional if its
removal helps detection on some attacks while hurting it on others.}
\label{tab:dissociation}
\resizebox{\columnwidth}{!}{%
\begin{tabular}{lccc}
\toprule
Heads & Attack-ID & Single-head EER & Conditional \\
\midrule
H0, H8 & 79--81\% & 11.3--11.7\% & no \\
H3, H4, H5, H10, H11 & 78--82\% & 11.4--17.9\% & yes \\
others & 76--79\% & 13.1--26.2\% & no \\
\bottomrule
\end{tabular}}
\end{table}

\vspace{-0.3cm}
\section{Discussion and conclusions}
\label{sec:discussion}
The heads of a single transformer layer adopt distinct roles: some carry the detection, others undermine it, and the rest are near-redundant. Two are directly usable. First, heads which work against detection can be removed outright: their zeroing lowers the EER with neither retraining nor additional parameters (13\% relative for WavLM, 22\% for Wav2Vec2). Second, the attention-entropy signature is a cheap, label-light diagnostic: its peak locates the most discriminative layer without training a probe, hence it can guide layer choice.

Our account is observational: we measure what each head contributes and where it attends, not how it computes. Single-head removal may understate redundancy, which the group sufficiency analysis mitigates. Our study spans two backbones and a single English corpus, ASVspoof~5, the single-head classifiers of Section~\ref{ssec:head-function} being applied to WavLM alone; other corpora and languages, and the same analysis for Wav2Vec2, are natural next steps. Even so, the message holds for both backbones: the attention head is an informative unit of analysis and opening the layer reveals structure, and usable gains, which a layer-level view misses.

\section{Acknowledgements}
This work was supported by the COMPROMIS project (ANR22-PECY-0011),
funded by the French Agence Nationale de la Recherche under the France
2030 program.

\section{Generative AI use disclosure}
   
Generative AI tools were used to support engineering tasks such as debugging analysis scripts and assisting with language translation. All research questions, design, analysis, and scientific claims were conceived and verified by the authors.

\bibliographystyle{IEEEtran}
\bibliography{references}

\end{document}